\DocumentMetadata{
  lang        = en,   
  pdfstandard = ua-2, 
  tagging     = on    
}
\documentclass[sigconf]{acmart-tagged}
\usepackage{graphicx}
\AtBeginDocument{%
  }

\copyrightyear{2026}
\acmYear{2026}
\setcopyright{cc}
\setcctype{by-nc-nd}
\acmConference[ASSETS '26]{The 28th International ACM SIGACCESS Conference on Computers and Accessibility}{October 25--28, 2026}{Vila Nova de Gaia, Portugal}
\acmBooktitle{The 28th International ACM SIGACCESS Conference on Computers and Accessibility (ASSETS '26), October 25--28, 2026, Vila Nova de Gaia, Portugal}
\acmDOI{10.1145/3797867.3829054}
\acmISBN{979-8-4007-2521-0/2026/10}

\begin{document}

\title[Generative-AI-Based Environment-Grounded VR Communication Training for Autistic Individuals]{From Grasping to Speaking: Generative AI-Based Environment-Grounded VR Communication Training for Autistic Individuals}

\author{Ziming Li}
\email{zl1398@rit.edu}
\affiliation{%
  \institution{School of Information,\\ Rochester Institute of Technology}
   \country{Rochester, NY, USA}
}

\author{Roshan L. Peiris}
\email{roshan.peiris@rit.edu}
\affiliation{%
  \institution{School of Information,\\ Rochester Institute of Technology}
   \country{Rochester, NY, USA}
}

\renewcommand{\shortauthors}{Li and Peiris}


\begin{abstract}

Autistic individuals often face barriers in workplace communication, where soft skills are embedded within ongoing tasks and surrounding environment context, not in isolated verbal exchange. Recent work has introduced LLM-driven agents into VR-based communication training and proposed prompting schemas that let agents generate dialogue grounded in the VR environment and the user's hand-based interactions. Building on this work, we explore how different levels of environmental grounding influence the training experience of autistic trainees and job coaches. We conducted an exploratory study with 9 autistic trainees and 7 job coaches across three modalities: conversation-only (C), conversation with environmental objects (C+O), and conversation with objects and grasp interactions (C+O+G). Usability and workload were comparable across modalities, while both trainees and coaches preferred the more interactive and environment-grounded condition (C+O+G). Participants viewed C+O+G as helpful for sustaining engagement and integrating communication practice into task performance. We discuss design considerations for more flexible, interactive, and workplace-relevant VR communication training for autistic individuals.

\end{abstract}




\maketitle

\section{Introduction}

Autistic individuals often face significant barriers when entering and sustaining employment, even when they demonstrate solid technical abilities and a strong motivation to work~\cite{nicholas_asd-vocational-support_2015,hedley_asd-intervention-literature_2017,Scott2019Autism,hendricks_employment_2010}. A major challenge involves job-related soft skills, including communicating with coworkers and customers, coordinating multiple tasks, and responding appropriately during fast-paced work routines. Vocational support programs and job coaches seek to address these challenges by providing behavioral support and individualized coaching to help autistic individuals develop job-related communication and task performance skills~\cite{Wehman2014Effect,WEHMAN2016Employment,vrPID}.

Workplace communication, however, rarely occurs as isolated verbal exchange. In real jobs, communication is embedded within ongoing task performance and coordinated action. Workers talk while handling objects, responding to requests, and referencing items in their physical surroundings. Thus, conversations are often shaped by what is currently being done, what has just happened, and what materials or tools are present. As a result, soft skills are highly situated, because many essential behaviors only emerge within the structure of real job tasks. For example, a trainee may need to notice a customer's cues while delivering items. Compared with community-based communication training, such as role play, which often abstracts communication away from concrete job materials and settings, workplace-based coaching allows trainees to practice communication as part of the workflow. Such in situ practice is important for helping autistic trainees generalize what they learn into stable job performance. 

However, workplace-based training is often difficult to access, hard to control, and challenging to repeat systematically. Opportunities are constrained by staffing, transportation, and business demands. Coaches cannot easily stage consistent repetitions of the same social situation, and real workplaces contain distractions and unpredictable events that limit targeted practice~\cite{LiJobCoach2025}. As a result, trainees may receive only limited and inconsistent opportunities to rehearse communication behaviors that are directly grounded in job tasks.

Technical simulation-based approaches, including virtual reality (VR), provide greater controllability, safety, and repeatability while simulating workplace environments. Here, VR can create immersive settings that preserve the sense of being on the job while reducing irrelevant disruptions~\cite{Parsons2002,Kandalaft2012,smith2014virtual}. However, many existing systems rely on pre-scripted dialogues or pre-programmed job tasks, limiting their ability to adapt to different work environments and individual trainee needs. Moreover, communication and hands-on action are often treated as separate components rather than an integrated workflow. This may reduce ecological validity and limit autistic trainees' opportunities to learn how communication is shaped in coordination with objects, surroundings, and task progress~\cite{bozgeyikli_vocational_2017,LiJobCoach2025}.

Recent advances in generative artificial intelligence (generative AI) introduce new possibilities for more flexible and context-responsive role play~\cite{brown2020language,shanahan2023role}. Large language model (LLM)-driven conversational agents can dynamically generate dialogue that adapts to scenario settings and user input in real time. Prior work has explored LLM-driven agents as rehearsal partners for conversation practice~\cite{Li2024LLMJobChatbot} and as tools to support everyday social problem solving for autistic individuals~\cite{Choi2024ChatGPTAutistic,Jang2024CHI}. Yet, even in systems that incorporate LLM-driven characters, training often centers on verbal exchange, with limited support for grounding dialogue in concrete object manipulation, their surrounding environments, and ongoing work activities. 

These gaps motivate the design of VR soft skills training platforms in which communication is embedded within a flow of work rather than practiced in isolation. Such platforms should support grounding dialogue practice in job-relevant actions, such as handing over, placing, or sorting items, while maintaining the controllability and safety of simulation. At the same time, richer interaction may introduce additional cognitive and sensory demands~\cite{Kourtesis2023-rv}. Given documented variability in sensory processing, attention, and executive functioning among autistic individuals~\cite{Kourtesis2023-rv,MARCO2011}, it remains an open question how different levels of environment grounding in generative-AI-based VR communication training affect the training experience, and whether richer environmental context and interactions enhance engagement and usability or increase complexity beyond trainees' comfort. Understanding both trainee and job coach experiences is therefore essential for designing flexible and accessible generative-AI-based environment-grounded VR communication training (hereafter, environment-grounded VR communication training) systems. Hence, in this work, we aim to address the following research questions:

\begin{quote}
\textbf{RQ1:} What are the usability and user experience of the two environment-grounded VR communication training modalities for autistic individuals: conversation with environmental objects (C+O) and conversation with environmental objects and grasp interactions (C+O+G)?

\textbf{RQ2:} How do autistic individuals and job coaches perceive the differences among conversation-only VR communication training (C), conversation with environmental objects (C+O), and conversation with environmental objects and grasp interactions (C+O+G)?

\end{quote}

To address these research questions, we conducted an exploratory study with 9 autistic trainees and 7 job coaches to investigate three VR interaction modalities of varying levels of environment grounding. It was in collaboration with a local vocational training institute that supports autistic people. The organization was already using VR and face-to-face role play alongside workplace-based training methods to develop trainees' soft skills. This paper presents the following contributions: (1) Empirical findings on how three interaction modalities (C, C+O, C+O+G) with increasing levels of environment grounding influence usability, training experience, and perceptions of autistic trainees and job coaches in environment-grounded VR communication training; and (2) Design considerations for implementing environment-grounded VR communication training for autistic individuals.

\section{Related Work}

\subsection{AI-Assisted Communication Training for Autistic Individuals}

Prior works have explored various digital approaches to support autistic individuals in practicing job-related communication skills. These approaches have primarily utilized virtual role-play, where virtual agents with predefined response logic or pre-recorded video clips simulate conversations, providing active and repetitive training experiences while reducing the anxiety of autistic individuals. The virtual role-plays are often delivered through computer screens~\cite{smith2014virtual,trepagnier2011virtual,Smith2020VITTAY,burke2018ViTA}, VR headsets~\cite{Adiani2022CIRVR,Li2024LLMJobChatbot,stanica2018vr}, or AR devices~\cite{Hartholt2019ARPoster}. For example, Smith et al. conducted a study where autistic individuals interacted with a virtual human resources representative across three difficulty levels, simulating different interview scenarios~\cite{smith2014virtual}. Further exploration has incorporated AI-driven virtual avatars, aiming to provide more flexible and adaptive conversation contexts~\cite{Adiani2022CIRVR, stanica2018vr}.

More recently, researchers have begun leveraging LLMs to generate more human-like and context-adaptive conversation simulation for autistic individuals~\cite{Choi2024ChatGPTAutistic,Li2024LLMJobChatbot}. These studies have identified both strong user preference for LLM-driven conversational agents as a supportive tool and concerns from practitioners about the quality and safety of the advice provided~\cite{Choi2024ChatGPTAutistic,Jang2024CHI}. Prior work has also noted ethical considerations in deploying LLM-driven agents with neurodivergent populations, including concerns about bias reinforcement and the perpetuation of stereotypes~\cite{papadopoulos2024LLMEthicalAutism,Lawrence2024-jh,Park2025Autistic}. 

\subsection{VR as a Training Medium for Autistic Users}

Previous research has widely explored VR as a training medium for autistic individuals, leveraging its controllable and immersive environments to support social and communication skills practice~\cite{Parsons2002,Newbutt2016,Masullo2025}. For instance, Kandalaft et al. developed the Virtual Reality Social Cognition Training program, in which autistic young adults practiced social interactions through avatars in a virtual environment~\cite{Kandalaft2012}. Follow-up studies further demonstrated improvements in social cognition and conversation skills among participants~\cite{Didehbani2016}. Kourtesis et al. developed VRESS, an immersive VR system for training social skills in autistic adults across five social scenarios with three difficulty levels, reporting positive outcomes in acceptability, system usability, and user experience~\cite{Kourtesis2023-rv}.

VR has also been applied specifically to workplace-oriented training for autistic individuals and individuals with intellectual disabilities. Bozgeyikli et al. developed VR4VR~\cite{bozgeyikli_vocational_2017}, which includes six transferable skill training modules designed for vocational rehabilitation of individuals with cognitive and physical disabilities, and evaluated its usability with autistic and neurotypical individuals. Babar et al. conducted a preliminary study on using VR as a job training tool for individuals with intellectual disabilities, identifying its potential and providing insights from the perspectives of job coaches~\cite{vrPID}. While these systems demonstrate the potential of VR for vocational training, the hands-on task components and communication training within them are typically designed as separate experiences. The integration of communication practice within a situated, task-driven context where the task state adapts to the trainee's actions remains underexplored.

\subsection{LLM-Driven Agents with Environmental Awareness}

Recent research has explored the use of LLM-driven agents across interactive environments. Park et al. developed a system of generative agents that can sustain coherent, memory-informed behavior across extended interactions in a simulated town environment, demonstrating human-like, believable social interactions and daily routines~\cite{park2023generative}. In the domain of robotics, prior work has proposed methods for grounding LLM agents in physical environments so that their reasoning reflects the surrounding scene and available affordances~\cite{vemprala2023chatgpt}, such as grounding language outputs in robotic affordances~\cite{ahn2022icanisay}, incorporating continuous environmental feedback into language-based planning~\cite{huang2022innermonologueembodiedreasoning}, and enabling open-ended exploration by perceiving and acting on game state~\cite{wang2023voyageropenendedembodiedagent}. 

To enhance the immersion of LLM-driven conversations, recent work has also explored integrating LLM agents with VR across domains such as communication practice~\cite{Li2024LLMJobChatbot}, language learning~\cite{pan2024ellmatembodiedllmagentsupporting}, and clinical training~\cite{zhu2025designingvrsimulationclinical}. Within VR, prior work has proposed environment-grounded prompting schemas that inject information such as environmental objects, user actions, and scenario state into the agent's reasoning, improving the contextual coherence of LLM-driven characters~\cite{LiEnvironmentAwareAgent2025}. However, these environment-grounded approaches have primarily been evaluated in non-training settings. Their application to communication training for autistic trainees and their job coaches remains unexplored. Thus, this paper aims to address this gap by investigating how progressively richer contextual grounding (C, C+O, C+O+G) influences the training experience.

\section{VR Training System Implementation}

The goal of this study is to provide a VR-based training system that simulates interactive and environment-grounded workplace communication scenarios. It enables trainees to engage in conversations with an environment-aware, LLM-driven avatar while interacting with relevant environmental objects. The VR training application is inspired by and built upon the system proposed by Li et al.~\cite{Li2024LLMJobChatbot}, and extends it to incorporate an environment-aware LLM-driven avatar~\cite{LiEnvironmentAwareAgent2025} which supports generating conversations based on the user's actions and surrounding VR environment.



To enable communication training that incorporates grasp interactions and environmental objects, we developed a VR application using Unity 2022.3.20f1 with Oculus Integration 57.0.2. The application 
runs on a Windows computer and is streamed to the headset via Meta Link. It uses OpenAI's GPT-4.1 (``gpt-4.1-2025-04-14'') model to generate human-like role-play responses based on several scenario prompt components, including scenario settings, object context, user action descriptions, and dialog history.

\begin{figure}[!h]
\includegraphics[width=\linewidth, alt={A four-panel figure shows a walkthrough of a fast-food VR training scenario from the trainee’s perspective. Panel (a) shows the start of the scenario, with the trainee standing behind a service counter facing a customer avatar in a restaurant interior. An instruction card appears in front of the trainee describing the scenario and task: serve the customer by talking based on the scene and surroundings, then pick up the ordered items and place them on the tray. A cash register and tray are visible on the counter. Panel (b) shows the conversation after the scenario begins. The customer avatar stands across the counter, and on-screen dialogue text shows the customer placing an order for a boxed burger and a coffee. Panel (c) shows the surrounding service area and available items. A drink machine, cups, and shelves containing boxed food items are visible, with selectable items highlighted. Panel (d) shows task completion in progress. The ordered coffee and boxed burger have been placed on the tray in front of the trainee, while the customer avatar remains across the counter. Additional on-screen text and a dictation prompt indicate the trainee can continue speaking with the customer while handling the order.}]{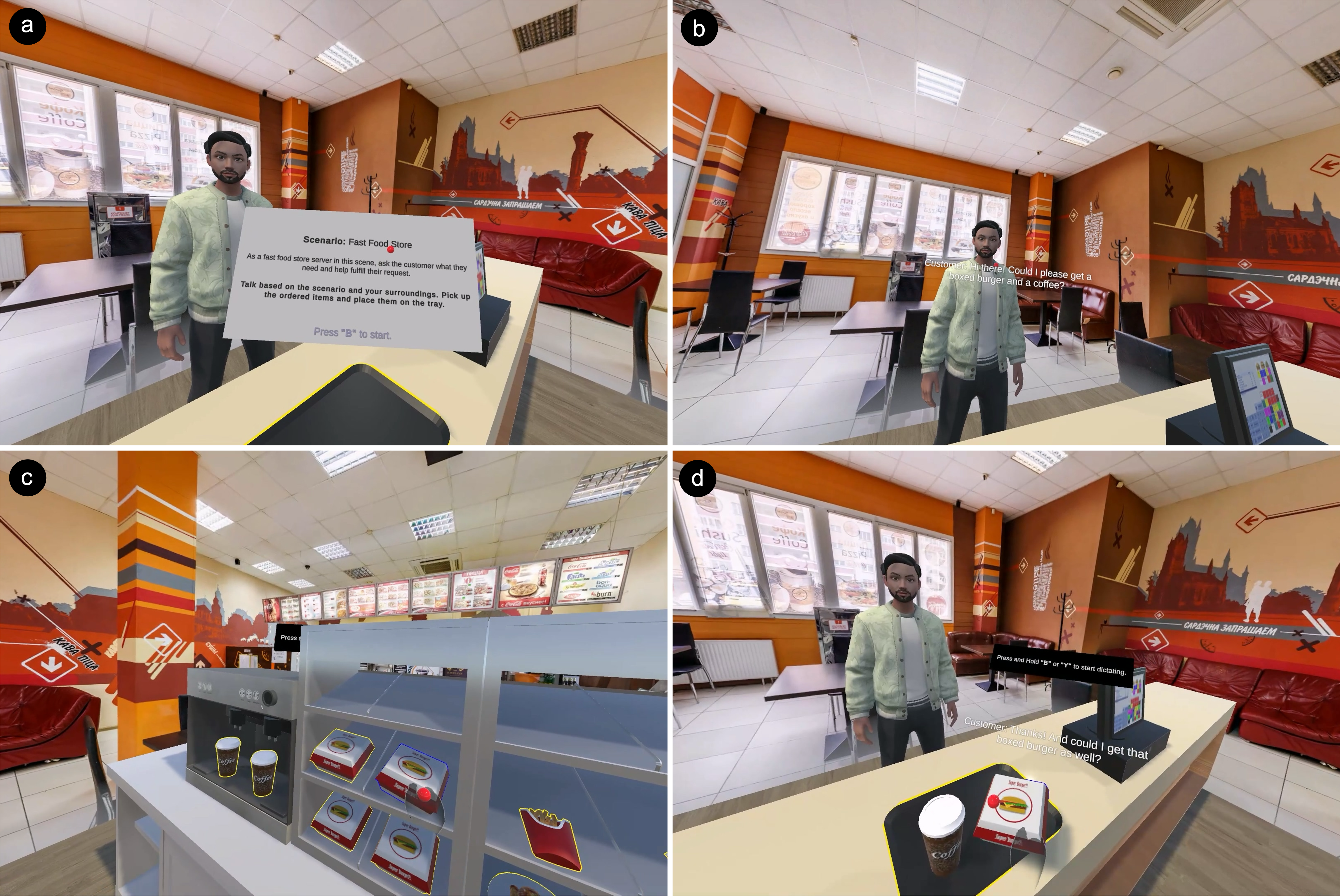}
  \caption{VR walkthrough of a training scenario, illustrated with the Fast-Food Restaurant scenario in the C+O+G modality. (a) Scenario instruction display; (b) Talking with the avatar; (c) Interacting with environmental objects based on the avatar’s request, shown here as grasping items from the back-area food preparation area, and (d) continuing the conversation with the avatar after placing the requested items on the tray in the front area. In the C and C+O modalities, the interaction in (c) and (d) is absent. The red sphere indicates the eye-tracking cursor, which was visible only on the mirrored desktop display and not in the VR headset.}
  \Description{A four-panel figure shows a walkthrough of a fast-food VR training scenario from the trainee’s perspective. Panel (a) shows the start of the scenario, with the trainee standing behind a service counter facing a customer avatar in a restaurant interior. An instruction card appears in front of the trainee describing the scenario and task: serve the customer by talking based on the scene and surroundings, then pick up the ordered items and place them on the tray. A cash register and tray are visible on the counter. Panel (b) shows the conversation after the scenario begins. The customer avatar stands across the counter, and on-screen dialogue text shows the customer placing an order for a boxed burger and a coffee. Panel (c) shows the surrounding service area and available items. A drink machine, cups, and shelves containing boxed food items are visible, with selectable items highlighted. Panel (d) shows task completion in progress. The ordered coffee and boxed burger have been placed on the tray in front of the trainee, while the customer avatar remains across the counter. Additional on-screen text and a dictation prompt indicate the trainee can continue speaking with the customer while handling the order.}
  
  \label{fig:vr_scenario_walkthrough}
\end{figure}

\subsection{Prompting Structure and Agent Communication}

To generate realistic role-play responses from the LLM that reflect environmental settings, user interactions, and ongoing conversations, we adapted a prompting structure based on the schema proposed in prior work~\cite{LiEnvironmentAwareAgent2025} for enabling environment-aware spatial interactions and conversational experiences. We simplified the original schema by removing the agent's action function list within the System Context component, the Spot component for agent's navigation, and agent's action validation feedback in the Communication component, as the virtual character in our application is designed solely to produce role-play speech responses without autonomous actions and movements. This design choice aligns with our goal of centering trainee interaction, reducing scene creation complexity, and improving responsiveness, an issue highlighted in prior evaluation regarding latency and limited naturalness in LLM-driven agent role-play action generation~\cite{LiEnvironmentAwareAgent2025}. 

Our adapted prompting structure consists of four main components: System Context (including system settings and scenario settings), Objects (including objects and containers), Characters (including character state and hand state), and Communication (including conversation messages and descriptive action cues, which are arranged chronologically). 
We further extended the schema by incorporating parent-object information in the user's grab and release cues (for example, ``Player's left hand grabs Coffee \textbf{from Counter}'') to provide richer environmental grounding and improve response accuracy. See Appendix A for the full system prompts we used in this study, and see Appendix B for an example conversation and action transcript between the user and the agent.

After the prompt is sent to the model facilitated by the user's action or speech, the model generates a response, which is then converted into speech using OpenAI Text-to-Speech (``tts-1'') API. Dialogues from the virtual character can optionally be displayed as captions, enabling job coaches who observe via external screen casting to more easily follow the conversation and provide guidance. Transcripts of conversations and user's actions during a session are recorded and stored on the device. 

\subsection{Scene Layout}

The training environment consists of a virtual character and a room-scale environment-grounded VR communication training environment. The virtual character is created using Ready Player Me character assets and features realistic facial animations. Lip movements are synchronized with generated speech using Oculus Lipsync, and eye movements are driven by the Realistic Eye Movements package to enhance social presence. The environment represents a virtual workplace with different layouts and interactable objects in the foreground. A 360-degree panoramic image captured from real workplace settings is rendered as the background to provide additional realism and immersion~\cite{Eiris2018,Ritter2021}. Within this room-scale space, trainees can move physically, interact with objects, and converse with the virtual avatar situated in the foreground. This combination of a room-scale interactive foreground and a 360-degree image backdrop is designed to enhance realism and spatial presence, while also simplifying the process of configuring training scenes that replicate real-world workplace settings.

The training environment in the foreground covers an activity area of approximately 3 meters by 2 meters and is structured into a front area and a back area to support lightweight representations of workplace layouts. Here, the virtual character is positioned in front of the front area to simulate a customer role.

Building upon the object classification used in prior work for describing grasp-based VR interactions and spatially-grounded conversations~\cite{LiEnvironmentAwareAgent2025}, we integrated scene items in three categories: Decorations, Objects, and Containers. Decorations are non-grabbable items used for set dressing (for example, a cash register) and represent ``fixed non-containers.'' Objects are grabbable items that can be picked up and relocated (for example, an apple), representing ``movable non-containers.'' Containers are non-grabbable items that hold Object items (for example, a basket), representing ``fixed containers,'' and allow Objects to be placed inside them, enabling hierarchical relationships such as coffee cups placed inside a coffee machine. Here, ``movable containers'' are excluded in our design to reduce interaction complexity and to keep spatial relationships predictable for trainees to support more consistent training experiences. In addition to these item types, structural elements of the front and back areas, such as cabinet or counters, are represented as ``fixed containers'' in the prompting structure because they can hold other objects within the scene layout. These classifications align with the Object components defined in the prompting schema~\cite{LiEnvironmentAwareAgent2025} and help the LLM-driven agent understand spatial relationships and generate environment-grounded responses during users' grasp interactions while relocating objects within the scene environment.

\subsection{User Interaction}

The training system supports grasp interactions through VR controllers, room-scale positional movement that reflects the trainee's real-world motion, and voice-based speech input. These interaction modalities aim to simulate realistic communication and general task behaviors commonly occurring in workplace environments. The interaction process is turn-based. After or before the virtual avatar completes its speech, it waits for the user to provide an action cue or initiate a conversation.

Trainees can interact with objects by pressing and holding either the hand trigger or index trigger on the left or right VR controller, corresponding to left or right hand grasping. They can also physically move within the scene to navigate the virtual workspace (see Figure~\ref{fig:vr_scenario_walkthrough}(c-d)). The application tracks user actions by recording their state, including hand occupancy and position, and incorporates this information into the context prompt. Each action is also transcribed into a descriptive text entry, which is provided to the virtual character as part of the conversational context when the trainee initiates dialogue. 

Trainees initiate speech by pressing a button on the controller. Speech input can be combined with other interactions, such as grasping and physical movement. Once initiated, the application records the speech and transcribes it using the OpenAI Speech-to-Text (``Whisper'') API. Then, the transcribed speech along with user's descriptive action cues is integrated into the context prompt and sent to the GPT model via the OpenAI API to initiate responses.

\section{Method}

To address the RQs, we conducted a user study with autistic trainees and job coaches to investigate the usability and user experience of VR role-play communication training that incorporates grasp interactions and environmental objects, as well as trainees' training experiences with these modalities.

\subsection{Study Design}

We conducted a within-subject exploratory study with autistic trainees and job coaches to investigate the training experience of different interaction modalities in role-play VR communication training. The three modality conditions were designed to incrementally introduce environmental and interaction complexity: (1) \textbf{Conversation-only (C)} serves as the baseline, replicating the interaction modality used in prior work~\cite{Li2024LLMJobChatbot}; (2) \textbf{Conversation with Environmental Objects (C+O)} adds visible environmental objects that both the trainee and the LLM-driven agent can reference in dialogue; and (3) \textbf{Conversation with Environmental Objects and Grasp Interactions (C+O+G)} further allows trainees to physically interact with objects through grasping. This incremental design allows us to examine how progressively richer environmental context and interaction capabilities affect usability, user experience, and training perceptions.

In this study, each trainee participant completed all three modality conditions, one per scenario. Because this is a within-subjects design, we used three structurally similar but contextually distinct scenarios (described in Section~\ref{sec:phase1_scenarios}) to reduce learning effects that could arise from repeating the same scenario across conditions. Both the assignment of scenarios to conditions and the order of conditions were randomized for each participant to mitigate ordering and carryover effects. Their job coaches observed the sessions and provided brief cues only if the trainee was unable to proceed. 

\subsubsection{Modality Conditions} 

The study compared three interaction modalities that varied in the level of environmental context and embodied interaction. All conditions featured the same LLM-driven virtual character and followed similar conversation goals to ensure comparability across modalities. 

\begin{figure}[!h]
\includegraphics[width=\linewidth, alt={A three-panel figure shows the same café training scenario across three interaction modalities from the trainee’s perspective. In all panels, the trainee stands behind a counter facing a customer avatar in a bright café interior with large windows. Panel (a) shows the conversation-only condition. The scene contains the customer avatar and dialogue text, but no visible interactive food or drink items are present on the counter. Panel (b) shows the conversation-with-environmental-objects condition. The customer avatar remains in front of the trainee, and pastries and cakes are now visible in a display case on the counter, with the items outlined to indicate selectable environmental objects. Panel (c) shows the conversation-with-environmental-objects-and-grasp-interactions condition. In addition to the visible pastry items, a drink and a slice of cake have been picked up and placed on a tray on the counter, showing direct object manipulation during the conversation.}]{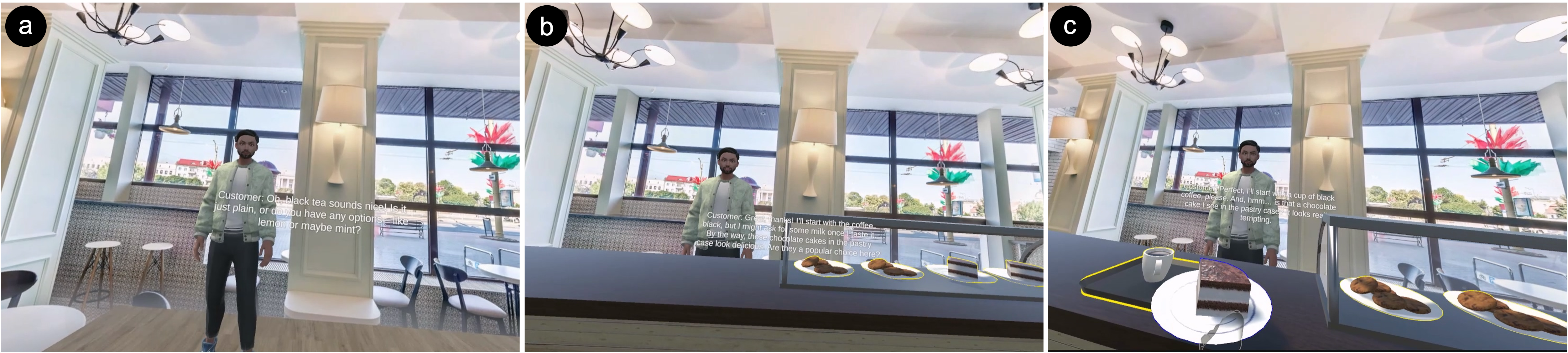}
  \caption{Three training interaction modalities evaluated in the study, illustrated with the Café scenario. (a) Conversation-only (C). (b) Conversation with Environmental Objects (C+O). (c) Conversation with Environmental Objects and Grasp Interactions (C+O+G), with a ``Tray'' object specifically implemented on the counter layout to evaluate physical task performance.}
  \Description{A three-panel figure shows the same café training scenario across three interaction modalities from the trainee’s perspective. In all panels, the trainee stands behind a counter facing a customer avatar in a bright café interior with large windows. Panel (a) shows the conversation-only condition. The scene contains the customer avatar and dialogue text, but no visible interactive food or drink items are present on the counter. Panel (b) shows the conversation-with-environmental-objects condition. The customer avatar remains in front of the trainee, and pastries and cakes are now visible in a display case on the counter, with the items outlined to indicate selectable environmental objects. Panel (c) shows the conversation-with-environmental-objects-and-grasp-interactions condition. In addition to the visible pastry items, a drink and a slice of cake have been picked up and placed on a tray on the counter, showing direct object manipulation during the conversation.}
  \label{fig:interaction_modalities}
\end{figure}

\begin{enumerate}
    \item \textbf{Conversation-only (C)}. In this baseline condition, participants engaged in verbal role-play with the virtual character without any environmental objects (see Figure~\ref{fig:interaction_modalities}(a)). The scene contained only a virtual character standing in front of the user with basic system context prompt that only describes the role-play settings, similar to the settings of the system used in prior work~\cite{Li2024LLMJobChatbot}, providing a conversational experience without spatial interaction or object-based context.
    
    \item \textbf{Conversation with Environmental Objects (C+O)}. Building on the C condition, this condition introduced environmental objects arranged within the scene layout and included a full prompting schema, which facilitated the generation of environment-aware responses of the LLM-driven agent (see Figure~\ref{fig:interaction_modalities}(b)). Participants could see the objects and reference them verbally, but could not grasp them. This condition enabled us to examine how adding workplace-relevant layouts and objects influenced participants' training experience and engagement.
    
    \item \textbf{Conversation with Environmental Objects and Grasp Interactions (C+O+G)}. Since grasp interactions require the presence of manipulable objects, the C+O+G condition built on C+O by further allowing participants to grab, hold, and release objects using VR controllers while conversing with the virtual character (see Figure~\ref{fig:interaction_modalities}(c)). A full prompting schema structure was included. This enabled trainees to perform object-based actions which were transcribed into descriptive cues and incorporated into the prompting structure for the LLM-driven character that facilitated environment-aware and spatial interaction-aware responses. Here, trainees were required to collect items ordered by the virtual avatar and place them into the tray to complete the order (see Figure~\ref{fig:vr_scenario_walkthrough}(a-d)). This condition enabled evaluation of how adding embodied interaction and physical task performance to the environmental context influenced communication experience and training engagement.
\end{enumerate}

\subsubsection{Scenarios} 
\label{sec:phase1_scenarios}

\begin{figure}[!h]
\includegraphics[width=\linewidth, alt={A six-panel figure shows three VR training scenarios from two viewpoints each. The top row presents the trainee's view, and the bottom row presents the corresponding avatar's view of the overall layout. Panels (a) and (d) show a butcher shop scene. In the trainee view, the user stands behind a glass display counter facing an avatar customer in front of a market-style background, with trays of meat and several small containers on the counter. In the avatar view, the full counter layout is visible from above and behind the avatar, showing highlighted food items arranged inside and behind the display. Panels (b) and (e) show a café scene. In the trainee view, the user stands behind a counter facing an avatar customer, with cakes or pastries displayed in a glass case and additional drink items placed farther back. In the avatar view, the broader service area is visible, including cakes in the front display and drinks and cups arranged on the back counter. Panels (c) and (f) show a fast-food restaurant scene. In the trainee view, the user stands behind a front counter facing an avatar customer, with a cash register and a tray on the counter against an orange restaurant interior. In the avatar view, the full service counter is shown from behind the avatar, including the register, drink machine, food shelves, and highlighted food and drink items arranged across the station.}]{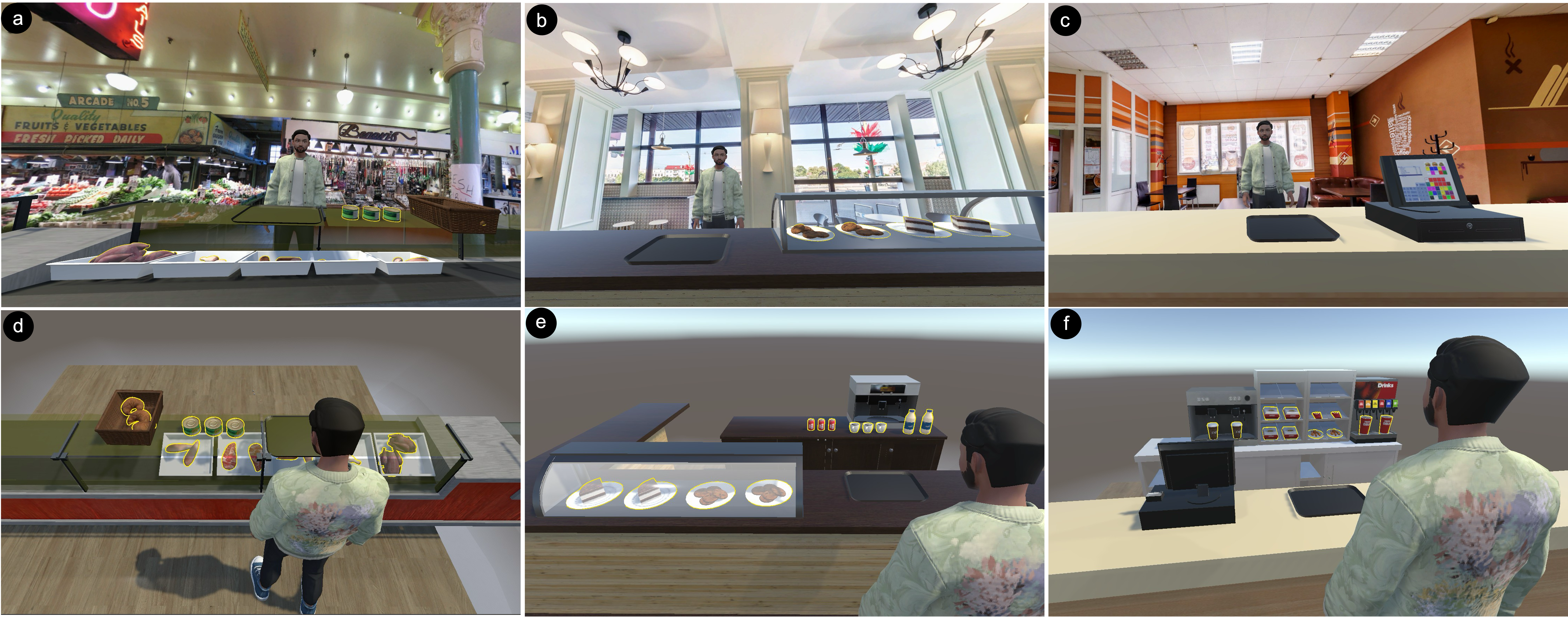}
  \caption{Three scenarios used in the study in the C+O+G modality, each shown from both the trainee’s and avatar’s perspectives. (a) Butcher Shop, trainee view. (b) Café, trainee view. (c) Fast-Food Restaurant, trainee view. (d) Butcher Shop, avatar view. (e) Café, avatar view. (f) Fast-Food Restaurant, avatar view.}
  \Description{A six-panel figure shows three VR training scenarios from two viewpoints each. The top row presents the trainee's view, and the bottom row presents the corresponding avatar's view of the overall layout. Panels (a) and (d) show a butcher shop scene. In the trainee view, the user stands behind a glass display counter facing an avatar customer in front of a market-style background, with trays of meat and several small containers on the counter. In the avatar view, the full counter layout is visible from above and behind the avatar, showing highlighted food items arranged inside and behind the display. Panels (b) and (e) show a café scene. In the trainee view, the user stands behind a counter facing an avatar customer, with cakes or pastries displayed in a glass case and additional drink items placed farther back. In the avatar view, the broader service area is visible, including cakes in the front display and drinks and cups arranged on the back counter. Panels (c) and (f) show a fast-food restaurant scene. In the trainee view, the user stands behind a front counter facing an avatar customer, with a cash register and a tray on the counter against an orange restaurant interior. In the avatar view, the full service counter is shown from behind the avatar, including the register, drink machine, food shelves, and highlighted food and drink items arranged across the station.}
  \label{fig:phase1_scenarios_preview}
\end{figure}

Based on discussions with job coaches, we designed 3 workplace-inspired role-play scenarios: (1) a Butcher Shop, (2) a Café, and (3) a Fast-Food Restaurant, each simulating common customer-order communication and task-handling workflows. All three scenarios are structurally similar (a customer ordering items from a server) to ensure comparability across modalities, while differing slightly in settings and interaction flow. These scenarios also followed the front-back layout and object-interaction structure commonly seen in customer-service workplaces. In every scenario, the trainee acts as a server and the LLM-driven virtual character plays the role of a customer. The trainee is expected to greet the customer, engage in small talk (such as commenting on the weather, the shop atmosphere, parking, etc.), and complete the customer's order based on the conversation. 

\begin{enumerate}
\item In the \textbf{Butcher Shop} scenario with environmental objects, all items are placed in the refrigerated counter in the front area (see Figure~\ref{fig:phase1_scenarios_preview}(a)(d)). Trainees complete the order by either verbally responding or physically identifying and handing over the requested items located in the front area.

\item In the \textbf{Café} scenario with environmental objects, items are distributed between the front counter and the cabinet in the back area (see Figure~\ref{fig:phase1_scenarios_preview}(b)(e)). Trainees complete the order by either verbally responding or physically selecting pastries from the front area and preparing drinks retrieved from the back area.

\item In the \textbf{Fast-Food Restaurant} scenario with environmental objects, all items are stored in the food preparation area at the back (see Figure~\ref{fig:phase1_scenarios_preview}(c)(f)). Trainees either verbally respond or physically navigate between the front and back areas to prepare and hand over the requested items during the conversation. 
\end{enumerate}

To create a more extended and realistic customer service interaction, the virtual customer is designed to place orders gradually rather than all at once. The customer may also request item swaps or adjustments, express indecisiveness, or show curiosity about certain items to enrich the back-and-forth dialogue. To promote consistent interaction flow across sessions, each scenario prompt also includes two example conversations as few-shot examples to guide the LLM-driven agent's behavior~\cite{brown2020language} (see Appendix A). Each scenario lasts approximately 5 to 10 minutes, depending on the flow of the conversation and the trainee's performance.

\subsection{Ethical Considerations}

The use of LLM-driven conversational agents in training contexts introduces potential risks, including the generation of contextually inappropriate or unexpectedly aggressive responses~\cite{papadopoulos2024LLMEthicalAutism,Park2025Autistic}. In this study, the LLM-driven agent was prompted to express a range of emotions, including satisfaction, annoyance, and hesitation, in order to simulate realistic customer interactions. This design decision was informed by discussions with job coaches prior to the study, who indicated that exposing trainees to varied customer temperaments was important for preparing trainees for real workplace situations. To mitigate the risks associated with this design, all training sessions were conducted under the supervision of job coaches, who could intervene or redirect the session if the agent's behavior became inappropriate or distressing. We also constrained the agent's behavior through system prompts that specified its role, scenario context, and communication boundaries (see Appendix A).

\subsection{Participants}

A total of 9 trainee participants and 7 job coach participants were recruited through a local non-profit job training organization that supports individuals with intellectual and developmental disabilities. The organization regularly uses role-play-based coaching methods to teach communication skills, conducts workplace-based training, and has been adopting VR technologies for both soft-skill and task-oriented training. Each trainee was paired with their assigned job coach from the organization's existing training program. All participants were identified and invited by the organization's manager to participate in the study.

The 9 trainees (labeled T01-T09), aged 21 - 39 (M = 29.56, SD = 5.68; 7 males, 1 female, and 1 non-binary), and the 7 job coaches, aged 24 - 60 (M = 38.43, SD = 12.51; 2 males, 4 females, and 1 non-binary), with 0.5 - 6 years of job coaching experience (M = 3.00, SD = 2.10), were selected based on availability within the organization's ongoing training schedules. Seven of the nine trainees reported having past work experience related to customer service, and eight had previously received conversation or social skills training. Eight trainees reported having used a VR headset before, with five of them using it approximately once a week and the remaining three having used it only a few times. All seven job coaches also reported having prior experience with VR headsets.

Here, J01 and J06 each supported two trainees: J01 supported T01 and T03, and J06 supported T06 and T09. Each of the remaining job coaches supported one trainee. For clarity in reporting, the job coach identifiers used in this paper are J01, J02, J04, J05, J06, J07, and J08. Each participant received 20 USD as compensation for participating in the study.

\subsection{Measures}

To address the two RQs, we employed a mixed-methods approach. For RQ1, we used the System Usability Scale (SUS)~\cite{brooke_1996_sus} to assess trainees' perceived usability, Raw NASA Task Load Index (NASA-TLX)~\cite{hart_1988_nasatlx,hart_2006_raw-nasatlx} to measure workload, and one 5-point Likert-scale item assessing trainees' overall preference for the interaction modality (``Overall, how much did you like this training interaction?''). The inclusion of SUS and NASA-TLX was motivated by the concern that richer environmental context and interactions may affect usability and introduce additional cognitive and sensory demands for autistic trainees~\cite{Kourtesis2023-rv}.

Job coaches completed a separate set of 12 structured rating items designed in consultation with them to capture dimensions they considered relevant for evaluating trainee performance and training experience in workplace communication scenarios, including learning objective achievement, conversation effectiveness, trainee engagement, perceived workplace resemblance, and willingness to use the system (see Appendix C for the full list). Some items were condition-specific, applicable only to modalities involving environmental objects and/or grasp interactions. All items used a 5-point Likert scale. Here, the eye-tracking indicator displayed on the laptop during sessions allowed job coaches to monitor trainees' gaze behavior as part of their observation, which is consistent with their existing practice in vocational communication training. Although eye contact was included in the job coach rating items as one dimension of the training evaluation, job coaches in this study applied these items flexibly based on their knowledge and experience of individual trainees' abilities and goals.

For RQ2, the moderator conducted session observations, noting trainees' interaction patterns, job coaches' coaching behaviors, and verbal comments made during the tasks. In-app session recordings and conversation transcripts served as additional data sources for the qualitative analysis. Post-study structured interviews were conducted separately with each trainee and job coach (or alternatively completed as a written survey, depending on preference) to understand how participants perceived the differences among the three modalities.

\subsection{Apparatus and Study Procedure}

The study was approved by the IRB office of the authors' institution. Trainee participants were screened for potential health risks (for example, eye strain, motion sickness, or vertigo) prior to participation to ensure their safety.

The study was conducted in a designated lab space at the collaborating job training organization, providing an open area of at least three meters by three meters suitable for room-scale VR activity. The equipment setup included a Meta Quest Pro headset, which supports eye tracking, a smartphone used to cast the VR view and in-app audio via the Meta Horizon app, and a Windows 11 laptop. The VR training application ran on the laptop and was streamed to the headset through Meta Link. This setup mirrored the headset's view on the laptop display and allowed job coaches to observe trainees' eye-tracking behavior through a moving dot indicator (visible only on the laptop screen). The laptop also stored experimental logs for later analysis.

Each study session involved one trainee participant and one job coach, who observed and supported the trainee throughout the procedure. The study consisted of a demographic survey, an experimental evaluation, and a post-study interview or questionnaire depending on participant preferences and availability. After a brief introduction and signing of informed consent forms, both trainee and job coach participants completed a demographic survey. 

Next, they proceeded to the experimental evaluation, which included a familiarization phase and a formal phase. During the familiarization phase, trainees engaged in a learning scenario using the C+O+G modality to become familiar with the VR controllers and system interactions, including grasping and relocating items and conversing with the virtual avatar. This phase lasted approximately 10 minutes. 

In the formal phase, the trainee experienced the three modality conditions in a counterbalanced order, with each condition randomly paired with one of the three formal scenarios. The job coach stood nearby and observed by listening to the in-app agent's speech through the smartphone casting, holding the phone to their ear to prevent any audio leakage that might distract the trainee, while monitoring the in-app view on the laptop screen with an eye-tracking cursor displayed that indicated trainees' eye-gaze during the tasks. During each session, the in-app view displayed on the laptop was recorded, as well as 
logs of conversations and interactions. The moderator also observed each session and took notes.

After each condition, the trainee completed a post-condition questionnaire and the job coach completed a separate post-condition questionnaire evaluating the trainee's performance and the overall training experience. A five-minute break was provided between conditions. 

Upon completing all conditions, a structured post-study interview was conducted separately with the trainee and the job coach (or alternatively completed as a written survey, depending on preference). The entire study lasted up to 120 minutes.

\section{Findings}

\subsection{Usability, Workload, and Preference}

We analyzed quantitative data collected from post-condition questionnaires completed by job coaches and trainees to examine the perceived usability, workload, preference, and training experience of each training modality.

\subsubsection{Trainee Ratings}

\begin{figure}[!h]
\includegraphics[width=\linewidth, alt={Three side-by-side bar charts compare trainee ratings across three conditions: C, C+O, and C+O+G, with error bars shown on each bar. Panel (a), titled "SUS Scores," shows mean usability scores increasing slightly across conditions, from 84.72 for C, to 88.61 for C+O, to 90 for C+O+G. Panel (b), titled "Raw NASA-TLX Scores," shows workload scores that are fairly similar across conditions, at 1.72 for C and 1.87 for both C+O and C+O+G, with the largest variability in C+O+G. Panel (c), titled "Overall Preference Rating," shows mean preference ratings increasing from about 4.44 for C, to about 4.67 for C+O, to about 4.89 for C+O+G.}]{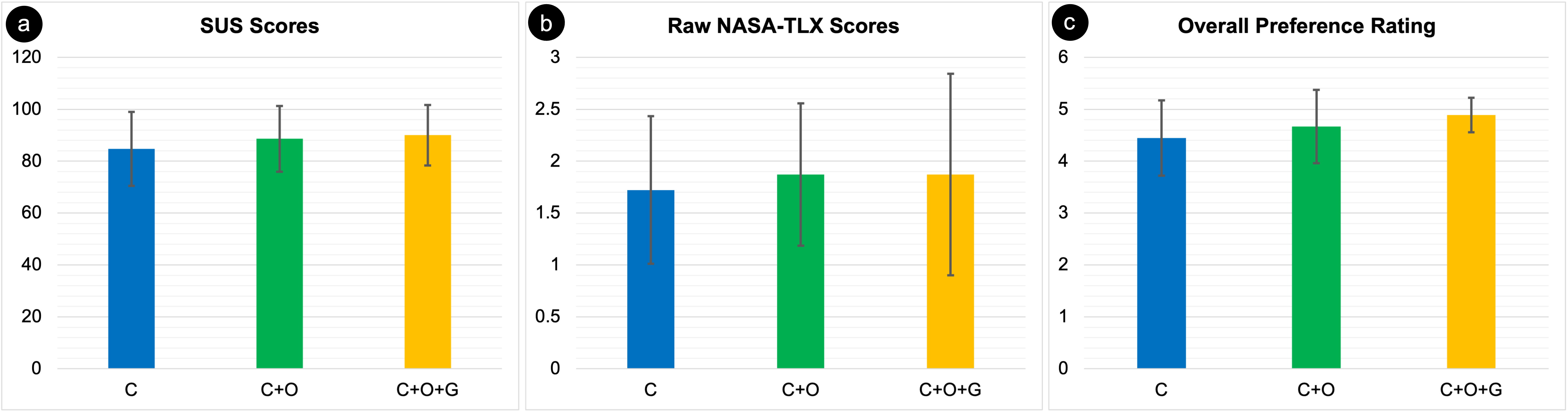}
  \caption{Trainees' ratings for the three conditions (C, C+O, C+O+G). (a) Mean SUS scores (0-100, derived from 5-point Likert items; higher is better). (b) Mean raw NASA-TLX scores (7-point scale; lower is better). (c) Mean overall preference ratings (5-point scale; higher is better). Error bars indicate standard deviations.}
  \label{fig:trainee_rating_graph_combo}
  \Description{Three side-by-side bar charts compare trainee ratings across three conditions: C, C+O, and C+O+G, with error bars shown on each bar. Panel (a), titled "SUS Scores," shows mean usability scores increasing slightly across conditions, from 84.72 for C, to 88.61 for C+O, to 90 for C+O+G. Panel (b), titled "Raw NASA-TLX Scores," shows workload scores that are fairly similar across conditions, at 1.72 for C and 1.87 for both C+O and C+O+G, with the largest variability in C+O+G. Panel (c), titled "Overall Preference Rating," shows mean preference ratings increasing from about 4.44 for C, to about 4.67 for C+O, to about 4.89 for C+O+G.}
\end{figure}

For trainee ratings, given the small sample size and repeated-measures design, we performed Friedman tests to compare SUS scores, Raw NASA-TLX scores, and overall preference ratings across the three conditions.

For perceived usability, the Friedman test showed no significant difference in SUS scores across conditions, $\chi^2(2) = 0.48$, $p = .786$, Kendall's $W = .03$. The mean SUS scores were 84.72 (SD = 14.28) for the C modality, 88.61 (SD = 12.69) for the C+O modality, and 90.00 (SD = 11.66) for the C+O+G modality (see Figure~\ref{fig:trainee_rating_graph_combo}(a)). These scores suggest consistently high perceived usability across all three conditions, with all mean SUS scores falling within the excellent range~\cite{bangor_sus-adj_2009}.

Similarly, no significant difference was found in Raw NASA-TLX scores across conditions, $\chi^2(2) = 1.27$, $p = .531$, Kendall's $W = .07$. The mean Raw NASA-TLX scores were 1.72 (SD = 0.71) for the C modality, 1.87 (SD = 0.69) for the C+O modality, and 1.87 (SD = 0.97) for the C+O+G modality (see Figure~\ref{fig:trainee_rating_graph_combo}(b)).

For trainees' overall preference rating, the Friedman test indicated a marginal effect of condition, $\chi^2(2) = 6.00$, $p = .050$, Kendall's $W = .33$. Given this borderline result, we did not further interpret pairwise differences. The mean preference ratings were 4.44 (SD = 0.73) for the C modality, 4.67 (SD = 0.71) for the C+O modality, and 4.89 (SD = 0.33) for the C+O+G modality (see Figure~\ref{fig:trainee_rating_graph_combo}(c)).

Overall, these results suggest that trainees experienced the three conditions similarly in terms of perceived usability and workload, with a marginal difference in overall preference across conditions.

\subsubsection{Job Coach Ratings}

For job coach ratings, since not all items were equally applicable across the three conditions (e.g., items related to environmental objects were only applicable in the C+O and C+O+G conditions; see Appendix C), we report these ratings descriptively instead of statistically comparing across conditions.

For the shared overall evaluation items, job coaches generally reported positive ratings across all three conditions. For example, ratings for whether the trainee achieved the learning objective were 4.78 (SD = 0.44) in the C modality, 4.44 (SD = 0.53) in the C+O modality, and 4.56 (SD = 0.53) in the C+O+G modality. Ratings for whether the trainee handled the conversation effectively were 4.56 (SD = 0.53) in the C modality, 4.33 (SD = 1.00) in the C+O modality, and 4.78 (SD = 0.44) in the C+O+G modality. 
Ratings for whether the trainee was able to maintain eye contact with the customer avatar during the session, as observed by job coaches via screen casting, were 4.44 (SD = 0.53) in the C modality, 4.56 (SD = 0.53) in the C+O modality, and 4.22 (SD = 0.44) in the C+O+G modality.
Job coaches also reported high willingness to use the interaction modality for future training, with mean ratings of 4.67 (SD = 0.71), 4.56 (SD = 0.73), 4.67 (SD = 0.71) for the C, C+O, and C+O+G modalities, respectively.


For condition-specific items, the ratings likewise suggested favorable perceptions of the trainee's interactions and realism of the scenarios. In the C+O modality, job coaches gave positive ratings across items related to object engagement, workplace resemblance in environment, and workplace resemblance in interaction (M range: 4.67-5.00). In the C+O+G modality, ratings were similarly positive across items related to object-handling ability, engagement, and workplace resemblance (M range: 4.67-4.78).

Overall, these descriptive ratings suggest generally positive job coach perceptions of the trainee's performance and training experience across all three conditions.

\subsection{Observed Task Performance Patterns}
\label{sec:finding:observed_task_performance_patterns}

During the experimental sessions, job coaches observed trainees performing tasks and were allowed to provide necessary guidance when trainees needed support in responding to the virtual customer. All trainees successfully completed their assigned scenarios across the three conditions to the satisfaction of their job coaches. While some trainees required occasional assistance, they were generally able to engage with the scenarios and work toward the training goals across conditions. 
To explore notable performance patterns during the sessions, we conducted a thematic analysis following Braun and Clarke's approach~\cite{Braun2006Thematic,clarke2017thematic}. The first author took observation notes during sessions, which together with in-task comments from trainees and job coaches and interaction transcripts served as data sources for the analysis. The first author directed the coding and theme development process. An AI tool (ChatGPT) was used to assist with organizing the observation notes. All codes were developed by the first author and verified against the original notes and session transcripts. Codes were then iteratively grouped into themes that captured notable patterns in trainees' task performance and job coaches' coaching practice across conditions.


\subsubsection{Environment-Grounded Interaction Encouraged Realistic Role-Play and Embodied Engagement}

Across the three modalities, we observed that trainees became progressively more engaged as environmental context and interaction capabilities increased. In the \textbf{C} condition, trainees relied primarily on verbal exchanges and imagination to sustain the role-play. T04 creatively named the butcher shop and discussed different cuts of meat despite having no visible objects to reference. Notably, T07 used hand gestures to pretend to hand over an order even though no interaction was available, suggesting a natural desire for embodied engagement that the C modality could not support. In the \textbf{C+O} condition, the presence of environmental objects provided visual anchors that some trainees leveraged during conversations, though not all trainees actively referenced them (discussed in Section~\ref{sec:finding:visual_and_environemtal_cues}). Meanwhile, T09 attempted to grab an object in the C+O condition but could not, as physical interaction was not enabled in this modality.

These patterns suggest that while environmental objects can provide helpful conversational context, the ability to physically interact with them further encourages engagement. This was most evident in the \textbf{C+O+G} modality. Here, all trainees were able to collect ordered items and place them into the tray based on the conversation context to complete the ordering and checkout process. Trainees also leveraged these interactions to make the scenarios feel more realistic. For example, in the Butcher Shop scenario, T02 held an item in each hand while suggesting options to the customer and returned the unordered one to the refrigerator. 
T02 commented that \textit{``It [the interaction with the environment] gives you an idea what happens in live time.''} Job coaches also used these interactions to support training. For instance, J06 guided T06 on how to respond if the wrong item was given to a customer, while J01 noted that \textit{``This modality [C+O+G] enables teaching our people multi-tasking skills.''} 
The environment-grounded interactions with LLM-driven avatars also allowed conversations to extend beyond the immediately presented content. 
For example, in the Café scenario in the C+O+G modality, T04 asked whether the customer would like cream in the coffee while grabbing a cup of hot coffee with his controller. 
These observations suggest that environment-grounded interactions can support flexible and engaging role-play during VR communication training.

\subsubsection{Visual and Environmental Cues Supported Conversations and Task Performance}
\label{sec:finding:visual_and_environemtal_cues}

Visual cues in the VR environment played an important role in supporting trainees' task performance. In the C modality, where no virtual objects were present, trainees were still able to maintain conversations with the virtual avatar, mainly relying on their creativity. When visual cues were available in the C+O and C+O+G modalities, many trainees actively used elements in the VR environment as contextual reference to support their responses. For example, in the Fast-Food Restaurant scenario in the C+O modality, T07 searched the food preparation area when the virtual customer asked about available food options. 

Compared to the C modality, the LLM-driven avatar's responses in the C+O and C+O+G modalities were constrained by the virtual objects defined in the schema, which guided trainees to focus their dialogue on the available items. For instance, when T07 suggested an item that was not present in the C+O modality, the virtual customer became confused, prompting the trainee to return to the available menu items. However, not all trainees actively relied on environmental cues. This was particularly evident in C+O, where some trainees (e.g., T05) paid little attention to the visible objects until prompted by the avatar's responses, possibly perceiving non-interactive objects as decorative elements.


On the other hand, several trainees, such as T04, T05 and T09, needed assistance with recalling orders and/or answering totals during the scenario-based conversations. This may have been related to limited visual support across conditions: trainees could not interact with objects in the C and C+O modalities, and the cash register in the C+O and C+O+G modalities did not display ordered items or running totals. Here, T03 also reported that he had to think harder about what to say in the C modality because there were no surrounding objects, noting that he would have more to talk about if objects were available as in the other conditions. These observations suggest that richer visual feedback could better support task performance and enable more realistic interactions.

\subsubsection{Job Coaches' Role in Environment-Grounded VR Communication Training}

Job coaches generally provided minimal guidance during the sessions rather than continuously intervening. According to the observation notes, coaches tended to step in only when trainees became stuck, missed important responses, or required social guidance. This suggests that the system allowed trainees to carry out most interactions independently while enabling job coaches to provide support related to task progression and socially appropriate responses when needed. For example, coaches supported task progression by prompting trainees to reflect on alternative responses (J02), guiding appropriate conversational behaviors such as greeting customers (J01, J06), and helping recall missed information (J01, J08). Coaches also occasionally reminded trainees about communication behaviors, such as maintaining eye contact (J07) or speaking louder (J06).





\subsection{Job Coach and Trainee Perspectives}


Following the experimental sessions, we collected feedback from trainees and their job coaches through either post-study questionnaires or interviews, depending on participants' preferences and availability. We then conducted a thematic analysis~\cite{Braun2006Thematic,clarke2017thematic} of the questionnaire responses and interview transcripts. The first author performed initial coding and iteratively grouped related codes into themes that captured recurring patterns in participants' feedback.

\subsubsection{The Role of Environmental Objects in VR Communication Training (C+O, C+O+G)}
\label{finding:role_of_objects}

Feedback from both job coaches and trainees suggested the environmental objects played an important role in VR communication training. Job coaches generally agreed that presenting environmental objects in the scene as contextual cues and references helped trainees develop their conversations. For example, J04 commented that \textit{``[The trainees] in many cases respond to the visual surroundings and it helps them with talking points with the [virtual] customers.''} Similarly, J02 noted that her trainee struggled in the C modality because \textit{``he didn't know, other than coffee, what he was selling,''} and suggested that visible items such as pastries on the counter could prompt the trainee to offer more to the customer. 
Several trainees' comments supported this perspective, noting the objects helped situate the conversation within a meaningful scenario context, with T07 stating that \textit{`` I felt [having environmental objects] is very easy than having to come up with stuffs on the flight.''} 
Participants also indicated that environmental objects contributed to the realism and immersiveness of the training experience. All trainees described the object-rich scenes as realistic, and job coaches (e.g., J06) noted that the added context and complexity helped trainees immerse themselves in the training.

However, job coaches also pointed out that the environmental objects could sometimes be confusing for some individuals, especially due to the relatively abstract item models and trainees' limited prior experience with certain workplace settings. For instance, J02 commented that \textit{``The butcher shop seemed confusing because T02 didn't know what items were in front of him,''} and J06 similarly noted that the objects could be ``a little bit overwhelming,'' explaining that some individuals in this population may be especially sensitive to whether virtual objects appear sufficiently realistic. To address this issue, J02 suggested that including signs or labels for environmental objects, similar to those found in real-world settings, could help interactions proceed more smoothly.

\subsubsection{The Role of Environment-Grounded LLM-Driven Responses in VR Communication Training}
\label{finding:role_of_llm_avatar_in_training}


Several trainees described their interactions with the virtual avatar as natural and human-like. For example, T01 commented, \textit{``I felt like I was talking to a real person and he [the virtual avatar] understood everything I said,''} and T03, T04, T07, and T09 expressed similar sentiments.

Job coaches also recognized the adaptive nature of the avatar's responses. 
J06 noted, \textit{``I was actually impressed. It seemed like it [the virtual avatar] didn't have a problem picking up, even when he [T09] would mispronounce things or stutter, it seemed to understand what he was saying. I think it's both good and bad.''} Here, J06 further suggested that the avatar could occasionally ask trainees to repeat themselves when they stutter or mispronounce words, which could make the conversation feel more natural. 
Meanwhile, job coaches emphasized that the avatar's real-time conversational responses contributed to meaningful training experiences. J02 described how this benefited one trainee who typically struggled with repetitive conversation patterns:

\begin{quote}
    \textit{``His [T02's] interactions with the customers were meaningful and on point. This can be difficult for [T02] since he can easily get caught in a circle, saying the same thing in multiple different ways, and becomes very difficult to redirect. The way the customer replied in real time made the training meaningful. It really felt like he [T02] was at work.''} (J02)
\end{quote}

However, participants noted that when the avatar asked multiple questions at once, it could overwhelm trainees. T03 mentioned, \textit{``In the Butcher [store scenario], [the avatar was] asking too many questions at once. [I] want it to ask one at a time.''} J06's observation echoed this concern, noting \textit{``Collecting items seemed very easy for the trainee [T06], talking and especially answering multi-part questions was more difficult.''} Although our prompt for the LLM-driven avatar specified that the avatar should order items gradually, it did not explicitly instruct the avatar to ask questions one at a time. This suggests that further constraining how the avatar structures its questions and responses could help make the interaction more accessible for autistic trainees.

Participants also noted that the avatar occasionally misinterpreted trainees' input. These issues were sometimes related to speech-to-text transcription errors or when trainees did not fully complete the press-to-speak interaction, which could cause frustration during the conversation training. 

\subsubsection{Combining Communication and Object Manipulation in VR Communication Training}

Job coaches noted that many trainees were able to quickly become familiar with and master the grasping interactions using VR controllers (J01, J02), and most trainees (7/9) 
did not consider the grasp interaction difficult. T07 mentioned that \textit{``I would say it [the grasp interaction] made it easier because you could physically see what is around you and interact with it to give it to the customer.''} T01 noted that the physical interaction initially made it somewhat more difficult for him to talk, but he was able to manage the interaction once he became more familiar with the controls. 

Job coaches also emphasized that incorporating object manipulation could help trainees practice multi-tasking skills. For example, J08 highlighted that the interactions could provide a more engaging training experience, stating, \textit{``This [the environment-grounded interaction] gives the trainee the opportunity to actively listen and stay engaged by completing a task such as placing order on food tray.''} 
J01 and J08 further suggested that such training could give trainees a sense of accomplishment when completing tasks, which would be important in real job settings. J02 also noted that the interactive and flexible nature of the task allowed trainees to correct mistakes during the process: \textit{``He [T02] liked the butcher shop because he could interact and pick up the different meats, although he didn't know what they were and got a little confused. He did correct himself when he made a mistake repeating the order to the customer.''}

However, some participants noted that multi-tasking could be challenging. J02 observed that \textit{``He [T02] stumbled a bit with his conversation, but I think that's because he was doing two things at one time.''} J01 similarly noted that \textit{``He [T03] garbling the products at first but he learned and did well.''} This challenge may have been related to the interaction design, which required trainees to press and hold the trigger button while also pressing and holding the dictation button on the VR controller. 
Job coaches noted that trainees could become more comfortable with multi-tasking in the VR communication training with additional practice, as J04 commented: \textit{``with practice I am sure he [T04] would learn to talk with the customer while handing the person items in the training exercise.''}

\subsubsection{Preferences for Different Interaction Modalities}

The C+O+G modality was most frequently preferred by both job coaches and trainees. Most job coaches (6/7) identified the C+O+G modality as one of the modalities they would most like to integrate into their regular training routines, citing its benefits in engagement, realism, and multi-tasking practice. 
All trainees reported that they liked the experience of the C+O+G modality, in which they could converse while collecting items, as it provided a more unique and in-person workplace training experience. 

Meanwhile, several participants noted the C+O and the C modalities also had their own use case, especially considering the diverse training scenario context and the diverse characteristics of the autistic population. For example, J02 preferred the C+O modality, stating, \textit{``I feel like it [the C+O modality] is a good starting point for everyone to get used to talking with people and learning about their environment and engaging in a longer more well rounded conversation.''} 
J08 suggested the C modality and C+O modality have their use case especially for those who had limited fine-motor skills: \textit{``Some participants may not have good fine motor skills, therefore may not be able to learn the hand sets, so they would benefit from the first conversation only training.''} T02 also echoed that all three modalities have their unique perspectives, noting, \textit{``each interaction are evenly good, in my opinion, each have their own aspect of individual variety and differentiates from each other.''}


\subsubsection{Benefits and Challenges of Environment-Grounded VR Communication Training for Autistic Individuals}
\label{finding:benefits_and_challenges}


Job coaches highlighted several advantages of the system for supporting job-related communication skills, describing it as providing a safe, manageable, and relatively distraction-free environment in which trainees could explore and practice communication in different workplace settings. They noted its potential to support meaningful conversation practice, build trainees' confidence, simulate interactions with unfamiliar people, and incorporate physical task-related actions into the training experience. For example, J02 emphasized that ``For people who don't think they have the ability to do a certain job, the VR training could boost their self esteem and willingness to step out of their comfort zone while still in a safe environment.'' J08 similarly commented:


\begin{quote}
    \textit{``This conversation training may be helpful for trainees whom have a hard time getting into the community, or have anxiety when working with others. This would help the trainee still have the opportunity to learn and practice conversation skills.''} (J08)
\end{quote}

J02 also pointed out that the avatar's real-time adaptive responses could help redirect trainees when their conversations became repetitive and circular, as discussed in Section~\ref{finding:role_of_llm_avatar_in_training}.

However, job coaches identified several challenges, including 
difficulty responding to multi-part questions, difficulty handling grasp and press-to-speak controls for trainees with limited fine motor skills, and distraction caused by the less photorealistic VR environment. In addition, J07 noted that the limited facial expressiveness and social presence of the LLM-driven avatar could affect the interaction, stating that ``there's no soul in it so everything is going be sort of robotic, which will probably throw the residents [trainees] off a bit.'' 
J06 observed that in traditional in-person role-play training, job coaches can more directly target specific behaviors that a trainee may be struggling with and can immediately redirect them when they say something inappropriate for a workplace context. In contrast, the VR training was described as simulating a more complete scenario, with greater emphasis on carrying the conversation through to completion. 

These comments suggest that job coaches may view VR-based and traditional communication training as complementary approaches depending on the training goal and the individual trainee's needs. Here, VR training offers a more complete and interactive workplace communication simulation, while in-person training may provide more socially nuanced interaction and more immediate opportunities for targeted support.

\subsubsection{Suggested Improvements for the VR Training System}

Participants suggested several improvements to the current VR training system. First, both trainees and job coaches expressed a desire for more interactive objects, more hands-on interactions, and a broader range of training scenarios to cover different workplace contexts (T09, J01, J02, J04, J06). Examples included mopping floors, stocking shelves, scanning objects, operating coffee machines and cash registers, and taking notes within VR. Because the current system primarily focused on grasping interactions, participants suggested that introducing more diverse and in-depth object interactions, especially with workplace equipment, could better support trainees in learning how to use equipment commonly found in real job settings while also increasing the realism and immersiveness of the training. 

Participants also suggested increasing the complexity of the virtual environment. As the current training scenes involved only a single customer in a back-and-forth interaction, adding more dynamic elements could make the experience feel more realistic. For example, participants suggested including background characters engaged in different activities (T02, T06, T07), additional customers waiting in line (J06), and environmental details that better reflected the conversation context, such as a rainy outdoor setting when mentioned by the customer (J06). Additionally, J06 and J08 emphasized that incorporating more distractions and environmental stimuli could help trainees better prepare for real workplace settings. As J08 explained:

\begin{quote}
    \textit{``The individuals I work with struggle to tune out other stimuli. For example, other customers talking, maybe a co-worker interrupting to ask a question, strong unpleasant odor, etc. I think it is important to think about other environmental stimuli may be noticed by trainees.''} (J08)
\end{quote}

In terms of social realism, participants such as T02, J01, and J08 suggested that virtual avatars could be designed with a wider range of appearances and personalities, including rude, upset, or standoffish behaviors, to help individuals prepare for more diverse and challenging customer interactions. 

Job coaches suggested adding signs or tooltips in the VR environment to provide clearer clues about interactive objects. For example, J02 mentioned that such supports could help trainees better understand available items, as discussed in Section~\ref{finding:role_of_objects}. Trainees also raised system responsiveness as an area for improvement. For instance, T09 noted that shortening the virtual avatar's response time (currently 3-5 seconds) would improve the experience.

Finally, participants highlighted the importance of making the training system more accessible and inclusive for a wider range of autistic trainees. J08 commented, \textit{``It would be nice if a wheel chair person could use the VR without staff having to move them. I have a participant that is not able to move her chair when her hands are on both the controllers.''} Given that the current training system relies on trainees physically moving within a room-scale VR environment and using controller-based interactions, participants suggested several ways to improve accessibility. These included allowing movement through joystick-based locomotion, adjusting the environment based on the user's eye-level height, as also noted by T09, and adapting the controls for individuals with limited fine motor skills (as discussed in Section~\ref{finding:benefits_and_challenges}).



\section{Discussion}

\subsection{Summary}

To investigate the usability and user experience of environment-grounded VR communication training for workplace conversation practice among autistic individuals, we developed an environment-grounded VR communication training system based on the architecture proposed in prior work~\cite{Li2024LLMJobChatbot}. This system extends the prior system by integrating an interactive VR foreground and an environment-aware LLM-driven avatar~\cite{LiEnvironmentAwareAgent2025} that generates conversations based on the user's grasping actions, movement, and surrounding VR context. 

Through an exploratory study with autistic trainees and job coaches, this work contributes findings on the usability and user experience of the three training modalities (conversation-only (C), conversation with environmental objects (C+O), and conversation with environmental objects and grasp interactions (C+O+G)) in environment-grounded VR communication training.

Our findings suggest that although the three training modalities showed similar perceived usability and user experience, C+O+G was preferred by most trainees and job coaches. Participants highlighted its benefits in keeping trainees engaged, encouraging them to talk more, and supporting multi-tasking by integrating physical interaction with conversation practice.

\subsection{Usability, User Experience, and Perceived Difference Across Environment-Grounded Training Modalities (RQ1, RQ2)}

To address RQ1 and RQ2, the three training modalities (C, C+O, and C+O+G) showed no significant differences in trainees' perceived usability (SUS) or task load (NASA-TLX), suggesting that the addition of environmental objects and grasp-based interactions did not introduce measurable usability barriers or additional workload for autistic trainees. This indicates that incorporating grasp-based interaction with the virtual environment into conversation practice did not noticeably compromise the core training experience.

Nonetheless, descriptive trainee preference ratings still trended slightly higher for more environment-grounded modalities, and the post-study interviews helped explain this pattern. Trainees and job coaches highlighted several benefits of the C+O+G modality: it kept trainees more engaged during practice, encouraged them to produce longer and more grounded conversational responses, and allowed them to draw on physical task skills that many trainees already possess. As job coaches noted, workplace communication often involves concurrent tasks. Employees typically talk while performing tasks such as stocking shelves or handling products, and by integrating grasp interactions and environmental objects with conversation practice, C+O+G enabled trainees to practice multi-tasking skills that real workplace settings demand. For trainees who are already comfortable with physical tasks but struggle with communication, the concurrent hands-on activity may also serve as an anchor that lowers the barrier to conversation~\cite{Kim_workplaceVR_2024}. At the same time, some trainees, particularly those without prior VR experience, reported initial difficulty coordinating controller operations in the C+O+G modality, such as grasping objects while simultaneously using press-to-speak. These challenges were observed and reported to diminish as trainees progressed through the training sessions.

Our findings also highlight that there is no one-size-fits-all modality for autistic trainees, which underscores the importance of designing training systems that can be adapted to different individual preferences and support needs. The autism spectrum encompasses a wide range of sensory, motor, and cognitive profiles~\cite{Ousley_autism_2014,Masi_overview-of-autism_2017}, and participants' responses reflected this diversity. For trainees with fine motor skill difficulties or those who are easily distracted by physical tasks, C+O may be more appropriate, as it provides the environmental context that supports situated practice without requiring direct object manipulation. Moreover, for scenarios where the training goal is purely conversational, such as small talk practice or job interview preparation, coaches noted that adding environmental objects may not always be necessary and could introduce unnecessary distraction.

While broadly preferred, the C+O+G modality also has current limitations that affect its applicability. Several participants noted that the range of supported interactions did not yet cover key workplace equipment such as coffee machines or cash registers, which limits the modality's ability to fully replicate certain workplace scenarios. This suggests that the training value of C+O+G is closely related to the fidelity and variety of the interactive objects available, and that expanding the interaction library is a priority for future development. We further discuss the design considerations of these findings in Section~\ref{sec:PartIV:design_considerations_environmental_objects}.

\subsection{Design Considerations}
\label{sec:PartIV:design_considerations}

\subsubsection{Design of Environmental Objects for Communication Training}
\label{sec:PartIV:design_considerations_environmental_objects}

Our findings suggest that environmental objects and task interactions in environment-grounded VR communication training serve primarily to ground conversation in realistic context and support trainee engagement, rather than to fully simulate workplace tasks. Coaches and trainees noted that having objects to reference and interact with encouraged trainees to say more and sustain longer conversations. This suggests that the design criterion for selecting objects should not be replicating everything present in a given workplace, but rather identifying which objects can provide meaningful conversational context. 

At the same time, coaches noted that objects should be visually recognizable and representative of their intended purpose, as overly abstract appearances may cause some autistic trainees to fixate on unexpected details. Adding contextual labels or visual indicators to objects, similar to price tags or signage in real workplaces, could help trainees quickly recognize items and stay focused on the communicative task.


\subsubsection{Step-by-Step Guidance for Structural Training}

Our findings revealed that while the system prompt instructed the LLM-driven agent to order items gradually, it did not explicitly constrain the agent from asking multiple questions within a single conversational turn. This led to instances where the agent combined several inquiries simultaneously, which overwhelmed some trainees. This observation highlights a design consideration for LLM-driven communication training that agent behavior should be structured at the level of individual conversational moves and also at the level of task flow. Future systems should ensure that the agent presents one request or question at a time and waits for the trainee to respond before proceeding. Visual cues in the VR environment, such as hints indicating the current step or relevant objects, could further support trainees in tracking the progress of the interaction and forming their responses. This step-by-step approach aligns with the principles of structured teaching~\cite{Mesibov2010}, which emphasizes breaking tasks into visually organized steps to reduce cognitive load and support independent performance in autistic individuals.

\subsubsection{Balancing Interaction Demands Based On Trainees' Needs}

While trainees and coaches overall preferred C+O+G mainly for its engagement benefits, our findings suggest that the C and C+O modalities also have practical value for different trainees and training goals. Future environment-grounded VR communication training systems should therefore allow coaches to configure the level of interaction complexity per session, such as enabling or disabling environmental objects and grasp interactions, so that the same system can serve trainees with varying preferences, abilities, and support needs, which aligns with the principle of ability-based design~\cite{wobbrock_ability-based-design_2011}.

\subsection{Limitations and Future Work}

This study has several limitations. First, this study included 9 trainees and 7 coaches from a single local vocational training institute. Recruiting participants for this study required identifying trainee-coach pairs who already work together, have workplace communication training needs, and were both willing and available to participate in VR-based training sessions, which constrained sample size. The trainees who participated in this study did not exhibit notable difficulties with fine motor skills or verbal communication, as they were already engaged in vocational training programs. As an exploratory study comparing three interaction modalities, our findings may not generalize to autistic trainees and coaches in other programs. Future work should involve larger and more diverse samples across multiple sites, and include trainees with a wider range of support needs to further examine how different ability profiles affect the training experience across modalities.


Second, as the study primarily aimed to investigate the usability and user experience of the system, it did not examine the extent to which communication skills practiced in VR transfer to real workplace settings or the long-term training effectiveness of the system. Future work should explore these questions over an extended period of use. For example, future studies could employ structured rubrics for job coaches to assess trainee communication performance across sessions. Future work could also explore using AI-driven analysis of conversation transcripts to assess metrics such as trainee's response relevance and conversational initiative.

Third, the conversation-only (C) condition in this study follows the interaction modality used in prior work~\cite{Li2024LLMJobChatbot}. The C+O and C+O+G conditions incorporate environmental objects and physical interactions into the dialogue, reflecting job coaches' interest in more visual and hands-on training for workplace communication~\cite{LiJobCoach2025}. Future work could explore other modalities for introducing environmental context, such as text-based or voice-based cues, to further investigate how different aspects contribute to the training experience.

Fourth, the current LLM-driven avatar had limited expressiveness and variety. The avatar remained stationary and its facial expressions could not convey emotions, which may have reduced the realism of the social interaction. Additionally, GPT-4.1's processing time for longer contextual prompts resulted in response delays of approximately 3 to 5 seconds, which could disrupt conversation flow. Future work could explore smaller or locally deployed models to reduce latency, as well as more diverse and expressive avatar designs to better approximate real-world social interactions.

Finally, scenarios in the current system were created by the research team, and coach-authored scenarios were not yet explored in this study. As coaches play a central role in conducting role-play-based training and tailoring training content to individual trainees, future work should develop authoring tools that allow job coaches, especially those without technical backgrounds, to create and refine environment-grounded training scenarios~\cite{LiJobCoach2025}, and study how they author and deploy these scenarios in practice through longitudinal deployments.

\section{Conclusion}

In this paper, we presented an exploratory study of three LLM-driven VR communication training modalities: conversation-only (C), conversation with environmental objects (C+O), and conversation with environmental objects and grasp interactions (C+O+G), with 9 autistic trainees and 7 job coaches at a local vocational training institute. Building on prior work that introduced LLM-driven agents and environment-grounded prompting schemas into VR communication training, we focused on how different levels of environmental grounding influence the training experience of autistic trainees and job coaches. Usability and workload were comparable across modalities, while both trainees and coaches preferred the more interactive and environment-grounded condition (C+O+G). Participants found the LLM-driven avatar natural and adaptive, and viewed the environment-grounded modalities, especially C+O+G, as helpful for sustaining engagement and embedding communication practice in task performance. Job coaches further described the system as a safe and controllable space that complements traditional role play and workplace-based coaching, while noting that appropriate interaction modalities also depend on individual abilities and training goals. The empirical findings and design considerations presented here contribute to the development of more flexible, interactive, and workplace-based VR communication training for autistic individuals.


\begin{acks}

We would like to thank Radhika Mandar Parkhi, a graduate student, for assisting with conducting three interview sessions during data collection. We would also like to thank the job coaches from The Arc Ontario for their interest, participation, and valuable feedback in this study. A special thanks to Andy Watson and Kathy Kanada from The Arc Ontario for organizing study sessions, facilitating communication with job coaches and trainees, and continuously providing support throughout the study process.

\end{acks}

\bibliographystyle{ACM-Reference-Format}
\bibliography{sample-base}


\end{document}